**Broadband Isolated Attosecond Pulses: Generation, Characterization, and Applications**

Michael Chini, Kun Zhao and Zenghu Chang

*Institute for the Frontier of Attosecond Science and Technology (iFAST),*
*CREOL and Department of Physics, University of Central Florida, Orlando, FL, USA*

**Abstract**

The generation of the shortest isolated attosecond pulses requires both broad spectral bandwidth and control of the spectral phase. Rapid progress has been made in both aspects, leading to the generation of the world-record-shortest 67 as light pulses in 2012, and broadband attosecond continua covering a wide range of extreme ultraviolet and soft x-ray wavelengths. Such pulses have been successfully applied in photoelectron/photoion spectroscopy and the recently developed attosecond transient absorption spectroscopy to study electron dynamics in matter. In this Review, we discuss the significant recent advancement in the generation, characterization, and application of ultrabroadband isolated attosecond pulses with spectral bandwidth comparable to the central frequency, which can in principle be compressed to a single optical cycle.

Since the first demonstration of isolated attosecond pulses from high-order harmonic generation in 2001[1], the evolution of time-resolved spectroscopy into the attosecond domain has already had a major impact on our understanding of ultrafast electronic processes in atoms, molecules, and condensed matter. Much of this progress has occurred in the last several years, with the proliferation of attosecond technology into many laboratories worldwide, as well as the maturation of attosecond spectroscopic techniques. Just as the extension of high power femtosecond laser pulses to the few-optical-cycle regime in the 1990's[2] opened new avenues in strong-field physics and led directly to the "attosecond revolution"[3], it is expected that development and application of shorter attosecond pulse sources may similarly inspire new breakthroughs in ultrafast science.

Few-cycle optical pulses require ultrabroad spectral bandwidths, comparable to or even larger than the central frequency, as well as accurate characterization and control over the spectral phase. Fully harnessing attosecond science will ideally allow for the extension of such few- or even single-cycle optical pulses from visible and infrared wavelengths, currently available from chirped pulse amplification (CPA) and optical parametric amplification (OPA), into the extreme-ultraviolet (XUV) and soft x-ray (SXR) spectral regions. Presently, however, the shortest achievable isolated attosecond pulses (67 $as^4$) are still significantly longer than those supported by the broadest attosecond supercontinuum spectra (Fourier transform limited pulse duration: 16 $as^5$). This discrepancy, detailed in Table 1, is due not only to the difficulty of generating ultrabroadband attosecond supercontinua, but also to the difficulty of characterizing and compensating the spectral phase over such a bandwidth and ultimately delivering the compressed pulse on a target for applications. The development of new techniques for generating, characterizing, and applying broadband isolated attosecond pulses is therefore crucial for the advancement of attosecond science.

In this Review, we describe the recent progress in the development of broadband isolated attosecond sources, supporting few- or single-cycle XUV pulses. Although such few-cycle pulses have already been demonstrated with central energies of 35 eV[6], 80 eV[7], and 90 eV[4], the technology is still relatively immature and new techniques are needed to make the next significant advances. This Review reports on the most recent progress in the generation, characterization, and application of broadband isolated attosecond pulses. In particular, we focus on those techniques which can be applied to *ultrabroadband* pulses, which are ideally compressible to a single optical cycle, and which hold significant promise for the extension of attosecond science to shorter pulse durations. The Review further covers the latest state-of-the-art in applications of broadband isolated attosecond pulses to the study of bound electron dynamics and to the measurement and control of correlated systems.

**Generating ultrabroadband attosecond pulses**

High-order harmonic generation (HHG) naturally results in the generation of attosecond pulse trains with ultrabroad spectral bandwidths in the XUV and SXR spectral regions, characterized by a relatively flat "plateau" spectrum extending from below target atom's ionization threshold[8, 9] to the high harmonic cutoff. The harmonic cutoff photon energy is proportional to the driving laser intensity $I$ before the atom is fully ionized and inversely proportional to the square of the driving laser wavelength $\lambda$, and can extend into the "water window" region of x-ray transmission[10, 11] when driven by Ti:Sapphire laser pulses ($\lambda$ = 800 nm) and beyond 1 keV with longer wavelength drivers[12, 13]. Generation of ultrabroadband attosecond pulses is therefore achievable only by making use of the full high-order harmonic spectrum, including both the plateau and cutoff harmonics. Isolated attosecond pulses from HHG have been experimentally demonstrated and characterized using a variety of experimental techniques (most of which have been covered extensively in previous reviews[3, 14-16]). These include spectral selection of half-cycle cutoffs[17-19] as in amplitude gating[7, 20, 21] and ionization gating[22-25], temporal gating techniques such as polarization gating (PG)[6, 26, 27] and double optical gating (DOG)[4, 28-30], and spatiotemporal gating with the attosecond lighthouse technique[31, 32].

Of these, only the PG, DOG, and attosecond lighthouse are suitable for the generation of ultrabroadband attosecond pulses, as illustrated in Figure 1. In spectral selection techniques, such as amplitude gating (Fig. 1a) and ionization gating (Fig. 1b), only the attosecond pulse produced by the most intense half-cycle of the laser (or the most intense half-cycle before ground state depletion, in the case of ionization gating) produces the cutoff spectral components with the highest photon energies (blue shading), whereas the low-energy spectral components (red shading) arise from multiple attosecond pulses produced by other half-cycles of the driving laser field with lower amplitude. Extracting an isolated attosecond pulse from the attosecond pulse train therefore requires that the highly modulated low-energy spectral components originating from interference of two or more attosecond pulses (Fig. 1e) are blocked by a suitable filter which transmits only the attosecond continuum spectrum at the cutoff. Such schemes preclude the generation of isolated attosecond pulses with ultrabroad bandwidths, since the usable cutoff spectrum (yellow shading) typically covers a relatively small portion of the total XUV/SXR spectrum. This limitation can in principle be overcome by the development of half-cycle near infrared (NIR) driving pulses through field synthesis of light transients[33, 34]. Currently, however, even the state-of-the-art laser pulses with 2.1 *fs* (~0.9 optical cycles) FWHM pulse duration[33] are too long to completely suppress the attosecond emission from multiple half-cycles. On the other hand, the PG, DOG (Fig. 1c), and attosecond lighthouse (Fig. 1d) techniques make use of both the plateau and cutoff spectral components, resulting in an ultrabroadband attosecond supercontinuum (Fig. 1f).

**Temporal gating.** Generation of broadband isolated attosecond pulses requires an attosecond "light switch" - a technique which can effectively turn on the high harmonic generation process during only a single half-cycle of the driving laser field, with the carrier-envelope phase (CEP) of the driving laser set so that only electron trajectories originating from a single ionization event

(electric field maximum) result in attosecond pulse generation. To date, the most effective switches are based on the polarization gating technique. Attosecond pulses can be generated efficiently when the driving laser field is linearly polarized, as explained by the semi-classical recollision model[35]. However, the conversion efficiency drops rapidly when the driving laser is elliptically or circularly polarized[26, 36, 37]. The yield of attosecond pulse generation decreases by ~50% for an ellipticity of only 0.1[27] and by about an order of magnitude for ellipticity of 0.2. Therefore, if a driving laser can be produced with linear polarization during only a single half-cycle, with all other portions elliptically polarized, then a single attosecond pulse can be isolated.

The driving pulse for polarization gating can be produced by combining two counter-rotating circularly polarized few-cycle laser pulses with a certain time delay between them[6, 27, 38, 39]. For a properly chosen delay[40], the resulting pulse can be linearly polarized only during the central half-cycle, with circularly polarized leading and trailing edges. The CEP dependence of the attosecond spectrum exhibits a $\pi$ periodicity, as one or two attosecond pulses can be emitted within the gate depending on the CEP. In 2009, ultrabroadband attosecond continua were generated using the polarization gating with spectral coverage between 60 and 140 eV, and supporting a transform limited pulse duration below 50 $as$[41], approaching the atomic unit of time. The primarily limitation of the polarization gating technique for the generation of even broader attosecond supercontinua by extending the spectral cutoff results from the reduction of the attosecond yield due to the depletion of the target atom ground state with increasing laser intensity. For the ~5 $fs$ driving lasers used in the experiments listed above, the required delay between the two circularly polarized pulses is comparable to the laser pulse duration, and the target atom can be fully depleted before the linear gate. Further reduction of the pulse duration would allow the use of higher driving laser intensities, but is extremely difficult.

This limitation of the polarization gating can be overcome with the addition of a linearly polarized second-harmonic field to the driving field with time-dependent ellipticity. In the asymmetric two-color field, attosecond pulse generation occurs only once per optical cycle[42, 43], instead of once every half-cycle, and the polarization gate can remain open for a full optical cycle. This technique is known as the double optical gating[28]. For the same input pulse duration, this means that the delay between the two elliptically polarized pulses can be reduced by a factor of two. This in turn reduces the probability that the ground state will be ionized before the gate, and allows higher laser intensities to be used. Since the ionization rate is highly nonlinear with respect to the driving laser intensity, the DOG technique allows a significant increase in the driving laser intensity when compared with polarization gating for the same pulse duration. The field asymmetry also results in a $2\pi$ periodicity of the attosecond pulse spectrum with the CEP. In 2009, attosecond supercontinuum spectra were generated from 8 *fs* driving lasers with an in-gate intensity of $1.4 \times 10^{16}$ W/cm$^2$ using the DOG technique, yielding a nearly 200 eV FWHM bandwidth supporting 16 *as* transform-limited pulses[5]. To date, this is the broadest attosecond supercontinuum spectrum which has been generated. By extending the DOG technique to the 5 *fs* pulses which are now available in many laboratories, DOG can be used with higher intensities, and it is expected that the cutoff could be further extended.

Another critical advance in the generation of ultrabroadband attosecond pulses has come with the development of the interferometric polarization gating[44] and generalized double optical gating[29] techniques, for which the driving laser pulse has elliptically (rather than circularly) polarized leading and trailing edges. Because the polarization gate width is directly proportional to the ellipticity $\varepsilon$ of the pulse, the use of elliptical (say, $\varepsilon = 0.5$) rather than circular ($\varepsilon = 1$) polarization can allow the generation of isolated attosecond pulses from longer driving lasers[29, 45] or reduce the ground state depletion when high intensity few-cycle pulses are used[46]. In addition to allowing further extension of the attosecond supercontinuum cutoff with higher intensity few-cycle driving lasers, the generalized double optical gating also hold promise for the development of *high flux* ultrabroadband attosecond pulses, when combined with high energy driving laser pulses[47].

**Spatiotemporal gating.** Another promising route to the generation of ultrabroadband isolated attosecond pulses is the attosecond lighthouse technique[31]. Whereas previous gating techniques have primarily made use of time-dependent laser waveforms, the attosecond lighthouse is the first to take advantage of spatiotemporal coupling in the driving laser field. In particular, the technique relies upon a driving laser field exhibiting wave-front rotation in order to generate angularly separated attosecond pulses. Normally, when a laser pulse is used to drive HHG, the attosecond pulses are emitted in a spatially confined beam, with divergence typically much smaller than the driving laser. On the other hand, when generated by a rotating wavefront, each attosecond pulse in the train will be emitted in a different direction, perpendicular to the instantaneous wavefront of the driving laser at the time of its generation. Therefore, if the wavefront rotation within one half-cycle of the driving laser field is larger than the divergence of the attosecond pulse, the attosecond lighthouse technique can result in the generation of an isolated attosecond pulse by spatial selection in the far field. The attosecond lighthouse technique is characterized by its unique dependence on the driving laser CEP – for different CEP values, the propagation direction of the attosecond pulse will change, as the attosecond pulse is generated at a different location in the rotating wavefront.

The attosecond lighthouse has been demonstrated experimentally in harmonics generated from a plasma mirror[48] and from a neon gas target[32]. In both cases, the isolated attosecond pulse is indicated by a spectrum which is continuous over its entire bandwidth and by the carrier-envelope phase dependence. In particular, for the gas-phase harmonics the continuum spectrum covers both the plateau and cutoff orders of the high-order harmonic spectrum, extending from 30 eV to more than 110 eV. Extension of the attosecond lighthouse technique to even broader bandwidths is relatively straightforward, as the divergence angle of the attosecond pulse tends to decrease with increasing photon energy, allowing cleaner selection of the isolated attosecond pulse.

**Extension to mid-infrared driving lasers.** In principle, every gating technique for the generation of ultrabroadband attosecond pulses could benefit from the use of few-cycle mid-infrared laser sources. This is primarily due to the well-known increase of the cutoff photon energy[12, 49, 50] with $\lambda^2$, allowing the generation of broader bandwidths than what is achievable with Ti:Sapphire. Additionally, recent studies have revealed a reduction of the attosecond frequency chirp[51] (proportional to $\lambda^{-1}$) and the improvement of the phase matching[13, 52, 53], which will be critical to the development of high flux attosecond x-ray sources. Recent improvements in the infrastructure for OPA and OPCPA with spectra between 1.3 and 4 μm[54-59] have led to several major advances in the generation of broadband high harmonic spectra[13, 60, 61]. The appearance of a continuum spectrum in such experiments is likely an artifact of the experiment, however, as the grating spectrometer has insufficient resolution to resolve the narrow spacing of high-order harmonics from mid-infrared lasers, and the dependence of the spectrum on the driving laser CEP has yet to be demonstrated.

**Characterizing Ultrabroadband Attosecond Pulses**

Characterization of broadband isolated attosecond pulses is also challenging. Nonlinear autocorrelation methods, which have been extremely successful in femtosecond science, cannot be straightforwardly extended to the attosecond domain, due to both the lack of effective nonlinear materials in the XUV/SXR spectral region and the lack of photon flux in the attosecond pulses. Instead, attosecond spectroscopy has relied upon pulse characterization methods which combine the attosecond pulse with a delayed NIR laser field. This forms the basis of the FROG-CRAB (frequency-resolved optical gating for complete reconstruction of attosecond bursts) technique[62], in which the attosecond spectral phase is encoded in laser-induced perturbations to the kinetic momentum spectrum of an electron replica of the attosecond pulse. Although the FROG-CRAB has been widely applied to the characterization of isolated attosecond pulses and attosecond pulse trains, it is not suited for the characterization of ultrabroadband attosecond pulses, as the reconstruction algorithm relies upon a narrow-bandwidth approximation[63], and alternative methods are needed.

**Phase Retrieval by Omega Oscillation Filtering.** In the FROG-CRAB, the combined action of the attosecond pulse and the delayed NIR dressing laser ionizes an atomic target, yielding a delay-dependent photoelectron energy spectrogram. Extracting the attosecond pulse duration from such a spectrogram is nontrivial, and requires approximate analytical descriptions of the laser-dressed photoionization process. In the case of FROG-CRAB, the assumption of narrow spectral bandwidth is made (known as the central momentum approximation in the literature), for which the spectrogram takes the mathematical form of a FROG trace, allowing the use of reliable FROG retrieval algorithms[64, 65]. For ultrabroadband attosecond pulses, however, a different approximation must be made in order to ensure the accurate retrieval of the spectral phase over the entire spectral bandwidth. For this reason, the PROOF (phase retrieval by omega oscillation filtering) was developed[63]. The PROOF technique, like the FROG-CRAB, relies upon the measurement of electrons photoionized by an attosecond pulse under the perturbation of a dressing laser field. In this case, the assumption is made that the dressing laser is weak, and can be treated using lowest-order perturbation theory, similar to the RABITT (reconstruction of attosecond beating by interference of two-photon transitions) technique for characterizing attosecond pulse trains[66].

When the interaction of the XUV/SXR and NIR lasers with the target atom is treated under the framework of second-order perturbation theory, the yield of photoelectrons with a fixed momentum $v$ oscillates with the time delay. The resulting photoelectron momentum spectrum $I(v,\tau)$ can then be written as a sum of three components: $I(v,\tau) = I_0(v) + I_\omega(v,\tau) + I_{2\omega}(v,\tau)$, as illustrated in Figure 2a. Here, $I_0$ represents the combined probability of one- (XUV/SXR) and two-photon (XUV/SXR+NIR) absorption, and does not depend on the delay. $I_\omega$ and $I_{2\omega}$ result from interferences between the different one- and two-photon pathways to a given final state, and oscillate with the delay $\tau$ at the NIR laser frequency $\omega$ and at $2\omega$, respectively. Because the attosecond pulse imprints its spectral phase on the photoelectron replica, the spectral phase differences between the interfering pathways are encoded in the time delay dependence of $I_\omega$, and can be extracted using a suitable algorithm[4, 63, 67, 68].

***In-situ* Characterization.** Temporal characterization of the attosecond emission can also be performed within the generation medium itself, by extension of an experimental technique pioneered by Dudovich, *et al.* for characterizing the emission time of different high-order harmonics in an attosecond pulse train[69]. In this *in situ* method, a weak second-harmonic field is added to the driving field for HHG with a variable time delay. Although the second-harmonic perturbation is relatively weak, it can result in measurable changes to the spatial[70] and spectral[42] character of the attosecond emission. Here, the addition of the second harmonic breaks the half-cycle symmetry of the driving laser field, resulting in the generation of both odd and even harmonic orders. The yield of a given even harmonic depends on both the sub-cycle time delay between the fundamental and second-harmonic fields and the harmonic emission time, and oscillates with the time delay between the two fields. Similarly to the PROOF technique, this oscillation can be interpreted as an interference arising from different quantum pathways resulting in the emission of the same even harmonic[71], and the emission times (spectral phases) of the different harmonics can be extracted from the phase of the oscillation. Such interferences are largely independent of the harmonic photon energy, and the technique can be extended to ultrabroad spectral bandwidths. Moreover, as an all-optical technique, the measurement time can be dramatically reduced as compared to PROOF or other photoelectron measurement techniques.

Until recently, the *in situ* technique relied on the measurement of discrete even harmonics, limiting its applicability to the measurement of the intrinsic frequency chirp of attosecond pulse trains[51, 71]. However, the technique has now been extended to the characterization of isolated attosecond pulses by taking advantage of spatio-temporal coupling of the two laser fields, resulting in measurable changes to the spatially- and spectrally-resolved attosecond pulse. When the weak second-harmonic field $E_p(t)$ is brought into the generation medium at a small angle $\theta_p$ with respect to the driving field $E_0(t)$, the spatio-temporal coupling of the two pulses results in a modification of the XUV/SXR wavefront and an angular deflection of the attosecond pulse's propagation direction[72], as illustrated in Figure 2b. Measurement of the spatial profile of the generated beam as a function of both the XUV/SXR photon energy and the time delay between the fundamental and second-harmonic pulses allows full space-time characterization of the amplitude and phase of the attosecond pulse in the generation medium, and, by extension, at all points in space. Although the *in situ* technique cannot be used to directly determine the temporal profile of the attosecond pulse when delivered to a secondary target, it can be used to accurately inform the design of attosecond compression techniques to ultimately deliver single-cycle attosecond pulses with ultrabroad bandwidths.

**Attosecond Pulse Compressors**

The primary goal in generating ultrabroadband spectra is to take advantage of the full spectral bandwidth, yielding ultrashort attosecond pulses at the single-optical-cycle level. Several techniques have been demonstrated recently which give promising results on this front, though there is still no "catch-all" technique to allow both ultrabroad bandwidths and complete compensation of the phase error. Here, we focus on those techniques which have been successfully implemented experimentally, though several theoretical proposals show promise for eliminating the intrinsic chirp of attosecond pulses in the generation medium through coherent field synthesis[73, 74].

**Foil filters.** The high harmonic generation process intrinsically imprints a positive frequency chirp on the emitted attosecond pulse, as the high-energy photons are emitted slightly after those with low energy. Within the plateau, this leads to a group delay which increases linearly with increasing photon energy. This positive chirp can be compensated by the negative group delay dispersion (GDD) of a suitably chosen foil filter[75, 76], as illustrated in Figures 3a and 3b. Unfortunately, this compression scheme is extremely limited – although foil filters with transmission windows and negative GDD in a variety of spectral regions are available, the transmission bandwidths are not tunable and do not approach the full bandwidth of the attosecond supercontinuum in general. Moreover, the overall transmission is low, and spectral phase distortions near the edges of the transmission windows tend to lengthen the attosecond pulse duration. However, foil filters are still the most successful method for compressing attosecond pulses, having first been used to demonstrate compression to a single cycle in 2006[6], and remain a staple of attosecond technology today.

**Chirped Multilayer Mirrors.** Another approach to the compression of attosecond pulses is the extension of chirped multilayer mirror technology from the visible/NIR spectral region to the XUV/SXR. Multilayer mirrors have long been implemented in high-order harmonic beamlines due to their relatively high reflectivity at normal incidence, and were applied as focusing optics in the first attosecond experiments[1]. However, despite early proposals to use chirped multilayer mirrors to compensate the spectral phase of high-order harmonics[77, 78], they were predominantly used in conjunction with metal foil filters, with the mirror serving only as a band pass filter, rather than aiding in the compression (as in the characterization of 80 *as* pulses in 2008[7]). The first chirped mirrors for attosecond pulses were realized in 2011[79], with reflectivity as high as 10% within a spectral bandwidth of 100-130 eV, supporting transform-limited pulse durations of 165 *as*. In the chirped mirror, different frequency components of the attosecond pulse penetrate to differing depths within the mirror before reflection, resulting in a difference in group delay for the different spectral components, as illustrated in Figure 3c and 3d. Although chirped multilayer mirrors are currently only available with limited spectral bandwidth, rapid progress is being made in the design, manufacture, and characterization technologies. Already, chirped multilayer mirrors capable of compressing 70 *as* pulses with spectrum between 300-350 eV[80] and sub-50 *as* pulses with spectrum between 20-112 eV[81] have been proposed, and it is likely that chirped multilayer mirrors supporting ultrabroad bandwidths may be available in the coming years as further technical advances are made.

**Single-cycle Attosecond Pulses**

Reliably compressing ultrabroadband attosecond pulses to the single-cycle limit in a wide range of wavelengths will require new technologies in attosecond science. Currently, the state-of-the art in pulse compression relies on combinations of multiple techniques. Generating transform-limited isolated attosecond pulses requires control over the high-energy spectrum and phase, where foil filters have positive GDD. This can be done by using phase matching to select only the spectral components which have flat phase. Attosecond pulses generated in a dense medium are known to exhibit a reduced cutoff, due to poor phase matching at the highest photon energies. Therefore, simply by tuning the pressure in the generation gas target, spectral selection can be used to optimize the attosecond pulse duration without changing the intrinsic chirp or filter properties. Using such phase-matched spectral selection, isolated 67 *as* pulses were produced using DOG[4]. While reduction of the generation gas pressure leads to a measurable broadening of the spectral bandwidth as the cutoff is increased, the pulse duration is expected to increase due to the positive filter GDD at higher energies.

**Application of Ultrabroadband Attosecond Pulses**

Isolated attosecond pulses hold great potential for time-resolved measurements on unprecedented timescales, which will only be realized with the development of high energy attosecond light sources[16] with ultrabroad bandwidths. Still, broadband isolated attosecond pulses have two major advantages over attosecond pulses generated from the high-order harmonic cutoff: The low photon energy plateau spectrum allows access to bound states of atoms and molecules, and ultrabroadband attosecond pulses can simultaneously cover multiple absorption edges in atoms, molecules, or complex materials. However, application of such pulses remains challenging, due to the lack of spectral and temporal resolution in traditional photoelectron spectroscopy. In this section, we discuss the recent experimental progress making use of novel techniques to obtain temporal and spectral resolution and to which ultrabroadband isolated attosecond pulses are directly applicable.

**Bound State Wavepacket Dynamics.** The spectrum of a broadband isolated attosecond pulse can simultaneously cover bound (or quasi-bound) and continuum states of an atom or molecule, resulting in the formation of a wavepacket consisting of both bound states and continuum states, which naturally evolves on attosecond timescales. A delayed NIR laser can then transfer population between bound states or ionize the excite states, promoting the bound electrons to the continuum. The wavepacket dynamics can be probed using attosecond electron interferometry[82, 83] or attosecond transient absorption spectroscopy[84, 85].

After the initial excitation of the target by the attosecond pulse, both the bound and continuum wavepackets evolve freely. The subsequent ionization of the excited states by the NIR laser results in a second, delayed continuum wavepacket, as illustrated schematically in Figure 4a.

As the time delay between the attosecond pulse and the NIR laser is varied, the interference of the two wavepackets is unveiled in delay-dependent oscillations of the electron spectrum. In analogy to quantum state holography[86], this interference can be classified as being between a reference (directly ionized by the attosecond pulse) and an object (delayed ionization of the bound wavepacket by the NIR laser) wavepacket, and the full complex amplitude, including the lifetime and phase, of each excited state in the bound wavepacket can be extracted. Such a measurement was recently performed in singly-excited states of helium atoms[82], but is applicable to more general systems as well.

Analogous experiments using transient absorption spectroscopy can similarly unveil the dynamics in bound wavepackets (Figure 4b). In transient absorption, instead of measuring the photoelectrons, the attosecond pulse is spectrally dispersed after interaction with the target. Again, scanning the time delay unveils interference oscillations in the transmitted spectrum[87-89], which are analogous to the quantum path interferences in the photoelectron measurements[90]. Additionally, the transient absorption experiments reveal several features which arise from bound-bound couplings and which are absent in the electron interferometry measurements, such as the effects of the instantaneous AC Stark shifts[91] and the formation of light-induced states[92]. In addition to the experiments in singly-excited states of helium, wavepacket interferences have also been observed in transient absorption of singly-excited neon[93] and in doubly-excited helium[94], and experiments are currently underway in molecules[95].

**Correlated Electron Motion.** In general, attosecond transient absorption appears to be a promising tool for the application of ultrabroadband attosecond pulses to the study of electron correlation in atoms[85, 94, 96] and more complex targets[97]. Most recently, attosecond transient absorption has been applied to the study of field-induced insulating-to-conducting state transitions in $SiO_2$ glass exposed to a strong laser field[98]. The experiments[99] revealed instantaneous and reversible field-induced modification of the insulating material, by probing changes in the transmission of the glass in the vicinity of the L-shell excitonic transition (~100 eV) of Si. The results have tremendous implications for the development of lightwave electronics[100], and indicate that transient absorption spectroscopy may allow first access to time-resolved processes and ultrafast switching in high-$T_C$ superconductors or semiconductor solar cells with ultrabroadband isolated attosecond pulses which can cover the absorption edges of all of the constituent atoms.

**Conclusion and Future Outlook**

Laser technology for the development of ultrabroadband attosecond light sources is rapidly maturing, and when combined with suitable attosecond pulse compression techniques, will soon lead to diverse sources of single-cycle attosecond pulses covering plateau and cutoff harmonics spanning the XUV and SXR spectral regions. Such pulses are ideally suited for applications in attosecond electron interferometry and attosecond transient absorption. By extending such attosecond sources and spectroscopic techniques to keV photon energies, the potential of attosecond technology to the study of electron correlation in complex materials can be realized. Finally, combining the techniques used to generate broadband attosecond pulses with those for generating high energy isolated attosecond pulses will allow for the generation of single-cycle pulses with gigawatt peak powers and petawatt peak intensities on target, which can be applied in true attosecond pump-attosecond probe experiments.

This work is funded by the DARPA PULSE program by a grant from AMRDEC, the National Science Foundation and the Army Research Office.

| Gating Method | Gas | Central Energy (eV) | FWHM Bandwidth (eV) | FT Limit (as) | Measured Duration (as) | Compression Method | Characterization Method | Refs |
|---|---|---|---|---|---|---|---|---|
| Amplitude Gating | Ne | 80 | 28 | 75 | 80 | Zr Foil | FROG-CRAB | 8 |
| Ionization Gating | Xe | 25 | 8 | 130 | 155 | Al Foil | FROG-CRAB | 26 |
| PG | Ne | 100 | 22 | 45 | 130* | Al Foil | FROG-CRAB | 7, 44 |
| DOG | Ne | 100 | 170 | 16 | 67 | Zr Foil, Phase Matching | PROOF | 4, 5 |
| Interferometric PG | Kr | 26 | 15 | 260 | N/A | None | N/A | 49 |
| Generalized DOG | He | 200 | 60 | 20 | 148* | Al Foil | FROG-CRAB | 30, 51 |
| Attosecond Lighthouse | Ne | 50 | 35 | 47 | N/A | None | N/A | 33 |

*The pulse durations for PG and GDOG were measured using attosecond pulses generated in Ar gas.*

**Table 1 | State-of-the art in ultrabroadband isolated attosecond pulse generation, compression, and characterization.**

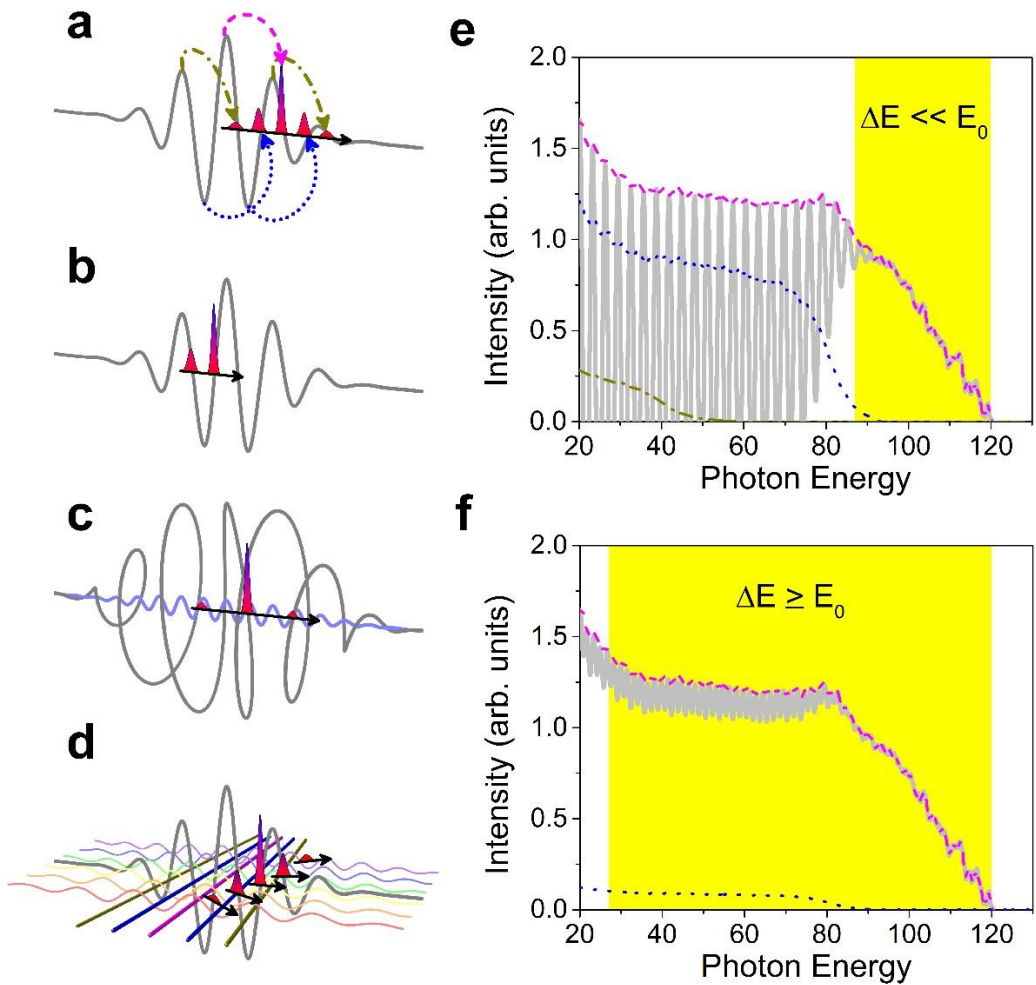

**Figure 1 | Gating an ultrabroadband isolated attosecond pulse.** Various techniques have been demonstrated for generating isolated attosecond pulses. **a,** Amplitude gating. **b,** Ionization gating. **c,** Polarization gating or double optical gating. **d,** Attosecond lighthouse technique. While only the cutoff spectrum is usable with amplitude gating or ionization gating (**e**), polarization gating, double optical gating, and the attosecond lighthouse technique are capable of producing ultrabroadband attosecond pulses with continuum spectrum covering the plateau and cutoff (**f**). The magenta, blue, and green lines indicate the contributions of different half-cycles to the total attosecond spectrum.

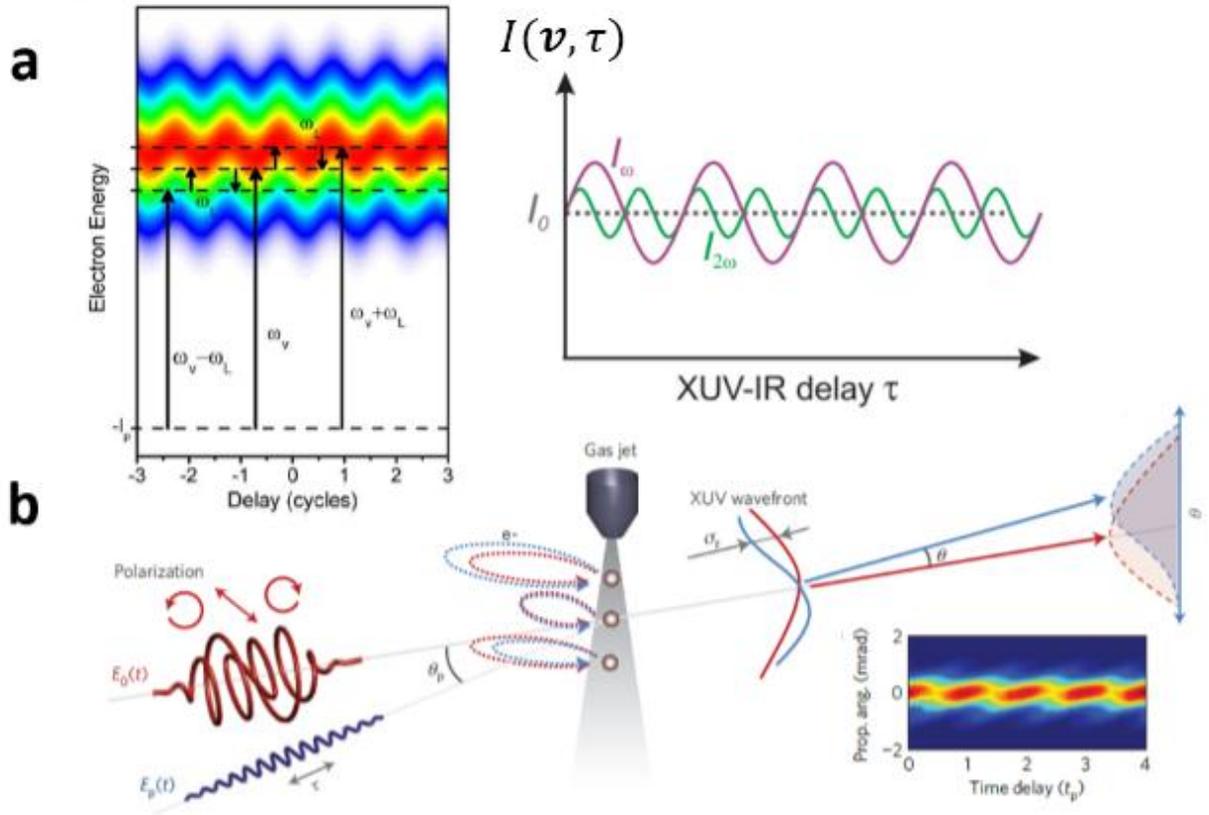

**Figure 2 | Attosecond pulse characterization.** Temporal measurements on ultrabroadband attosecond pulses requires methods which do not require narrow-bandwidth approximations. **a,** In PROOF, the weak perturbation of the dressing NIR laser field imprints an interference oscillation on measured delay-dependent photoelectron momentum spectrum (Figures adapted from Chini *et al.*[63] and Laurent *et al.*[68]). **b,** *In situ* characterization relies upon a weak perturbation to the attosecond pulse generation, resulting in an angular deflection of the attosecond beam which can be measured in the far field (Figures adapted from Kim *et al.*[72]).

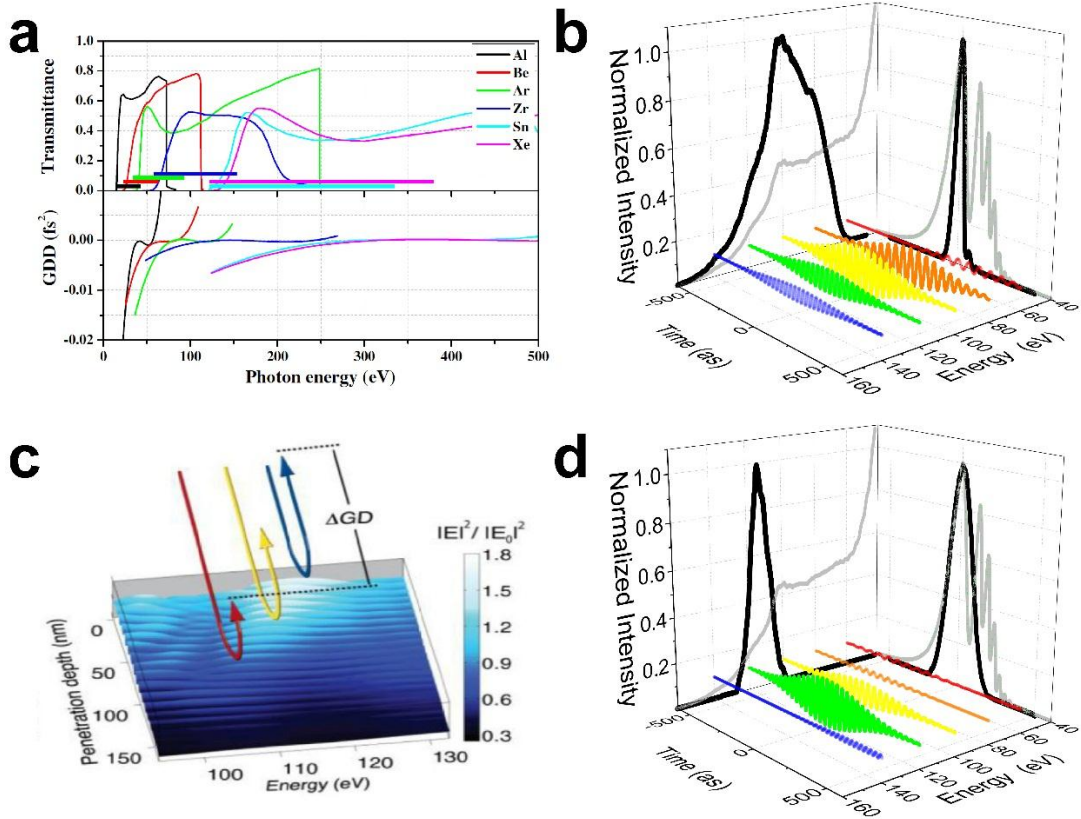

**Figure 3 | Conventional methods for attosecond pulse compression. a,** Various metal foil filters have relatively high transmission and negative GDD over a relatively broad energy range (Figure adapted from Ko *et al.*[76]). **b,** Spectrum and temporal profile before (grey lines in projections) and after (black lines in projections) passing through the Zr foil filter. The attosecond chirp is compensated so that different frequency components (colored lines in 3D) of the attosecond pulse add in phase to generate a short pulse. **c,** In a chirped multilayer mirror, different frequency components of the attosecond pulse (colored lines) penetrate to different depths within the mirror, imprinting a change in the group delay across the reflected spectrum (Figure adapted from Hofstetter *et al.*[79]). **d,** Chirped mirrors currently have limited bandwidth, but are more flexible than metal foil filters.

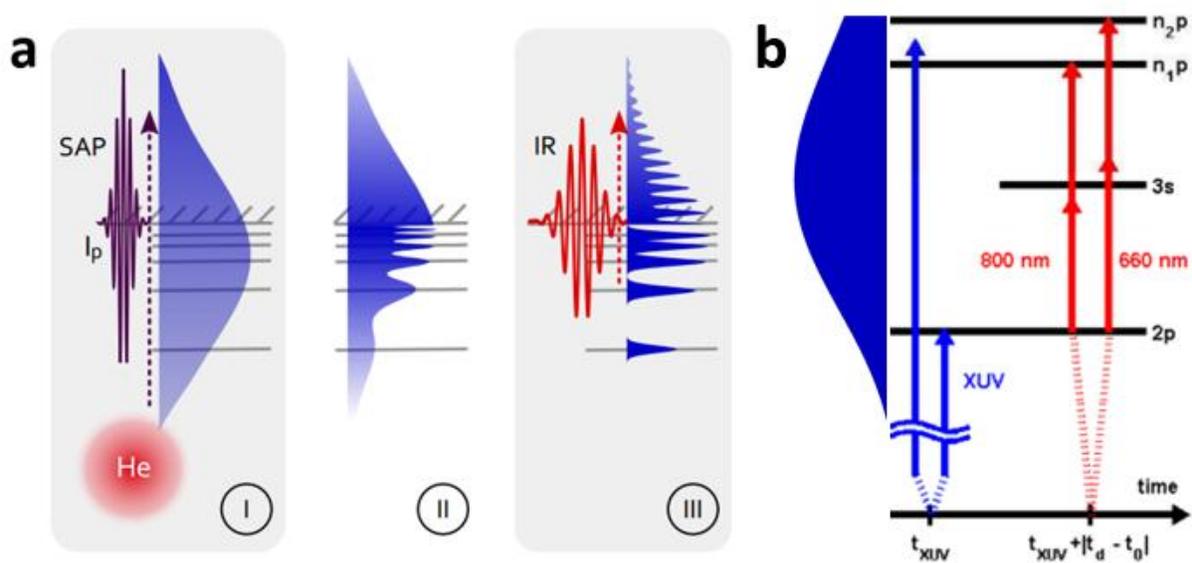

**Figure 4 | Schematic of attosecond electron interferometry and attosecond transient absorption measurements in helium. a,** In attosecond electron interferometry, an isolated attosecond pulse excites the helium atom to both bound and continuum states (I), which freely evolve in time (II) until the arrival of a delayed NIR laser (III), which ionizes the bound excited states. Interferences between the two ionization pathways are measured in the photoelectron spectrum (Figure adapted from Mauritsson *et al.*[82]). **b,** In attosecond transient absorption, additional features can be observed due to bound-bound coupling, for example between the 1s2p, 1s3s, and 1s$n$p states (Figure adapted from Chen *et al.*[90]).